\documentstyle[12pt]{article}
\newcommand{\ptra}{$p_{\mathrm{T}}$}
\newcommand{\mtra}{$M_{\mathrm{T}}$}
\newcommand {\omutau} {$\nu_\mu \rightarrow \nu_\tau$}
\newcommand {\omue} {$\nu_\mu \rightarrow \nu_e$}

\newcommand {\omuebar} {$\bar{\nu}_\mu \rightarrow \bar{\nu}_e$}
\newcommand {\oemubar} {$\bar{\nu}_e \rightarrow \bar{\nu}_\mu$}
\newcommand {\oetaubar} {$\bar{\nu}_e \rightarrow \bar{\nu}_\tau$}

\newcommand {\xedec} {$\tau^- \rightarrow e^-\nu\bar{\nu}$}
\newcommand {\xmudec} {$\tau^- \rightarrow \mu^-\nu\bar{\nu}$}
\newcommand {\xpidec} {$\tau^- \rightarrow \pi^-\nu$}
\newcommand {\xrhodec} {$\tau^- \rightarrow \rho^-\nu$}
\newcommand {\zrhodec} {$\tau^- \rightarrow \pi^- \pi^0 \nu$}
\def\xa1dec{$\tau^- \rightarrow a_1^-\nu$}
\def\za1dec{$\tau^- \rightarrow \pi^- \pi^+ \pi^- \nu$}
\def\yrho_2pi{$\rho^- \rightarrow \pi^-\pi^0$}
\def\ya1_3pi{$a_1^- \rightarrow \pi^-\pi^+\pi^-$}
\def\ypi_gg{$\pi^0 \rightarrow \gamma\gamma$}
\begin {document}
\title{Emulsion Chamber with Big Radiation Length 
for Detecting Neutrino Oscillations}
\author{A.E. Asratyan\thanks{Corresponding author. Tel.: (095)-237 0079. 
E-mail address asratyan@vxitep.itep.ru.},
G.V. Davidenko, A.G. Dolgolenko, V.S. Kaftanov, \\
M.A. Kubantsev\thanks{Now at Kansas State University, Manhattan}, 
and V.S. Verebryusov \\
\normalsize {\it Institute of Theoretical and Experimental Physics,}
\normalsize {\it Moscow, 117259 Russia}
}                         
\date {\today}
\maketitle

\begin{abstract}
A conceptual scheme of a hybrid-emulsion spectrometer for investigating
various channels of neutrino oscillations is proposed. The design emphasizes
detection of $\tau$ leptons by detached vertices, reliable identification 
of electrons, and good spectrometry for all charged particles and photons. 
A distributed target is formed by layers of low-Z material, 
emulsion-plastic-emulsion sheets, and air gaps in which $\tau$ decays are 
detected. The tracks of charged secondaries, including electrons, are 
momentum-analyzed by curvature in magnetic field using hits in successive 
thin layers of emulsion. The $\tau$ leptons are efficiently detected in
all major decay channels, including \xedec. Performance of a model
spectrometer, that contains 3 tons of nuclear emulsion and 20 tons of
passive material, is estimated for different experimental environments.
When irradiated by the $\nu_\mu$ beam of a proton accelerator over a
medium baseline of $\langle L/E_\nu \rangle \sim 1$ km/GeV, the spectrometer
will efficiently detect either the \omutau\ and \omue\ transitions in the 
mass-difference region of $\Delta m^2 \sim 1$ eV$^2$, as suggested by the 
results of LSND. When exposed to the neutrino beam of a muon storage ring
over a long baseline of $\langle L/E_\nu \rangle \sim$ 10--20 km/GeV, the
model detector will efficiently probe the entire pattern of neutrino 
oscillations in the region $\Delta m^2 \sim 10^{-2}$--10$^{-3}$ eV$^2$, as 
suggested by the data on atmospheric neutrinos.
\end{abstract}
PACS numbers: 14.60.Pq, 14.60.Fg
\\Keywords: neutrino oscillations, $\tau$ leptons, nuclear emulsion

\newpage
\section{Introduction}

     Oscillatory transitions among neutrinos of different flavors have 
emerged as a major topic of particle physics \cite{review}. An accelerator 
experiment using muon antineutrinos from $\mu^+$ decays at rest over an 
effective baseline of  $L/E \sim 1$ m/MeV, LSND at Los Alamos, has reported a 
positive signal in the channel \omuebar\ with a probability of
$\sim 3 \times 10^{-3}$ \cite{lsnd}. A consistent, albeit less significant,
signal was also observed in the CP-conjugate channel \omue\ using muon 
neutrinos from $\pi^+$ decay in flight \cite{conjugate}. When combined with 
the upper limits imposed by other accelerator and reactor experiments 
\cite{bugey, e776}, the LSND data suggest that the \omuebar\ oscillation is 
driven by a mass difference squared, $\Delta m^2$, between some 0.3 and 2.3 
eV$^2$. The transitions \omuebar\ and \omue\ in this region of $\Delta m^2$ 
will soon be explored with higher sensitivity by the BooNE experiment at 
Fermilab \cite{boone}.

     Qualitatively, an analogy with the quark sector suggests that (i) 
the $\nu_\tau$ should pick the largest contribution of the heaviest mass
state $\nu_3$, and that (ii) the mixings between the neighboring neutrino 
flavors ({\it i.e.}, $\nu_e \leftrightarrow \nu_\mu$ and 
$\nu_\mu \leftrightarrow \nu_\tau$) should be the strongest. Therefore, a
mass difference as large as $\Delta m^2 \sim 1$ eV$^2$ should primarily 
manifest itself in the $\nu_\mu \rightarrow \nu_\tau$ channel. 
The existing upper limits on the effective mixing $\sin^2 2\theta$ for 
the transition \omutau\ \cite{cdhs, e531, cern} largely come from 
experiments with small effective baselines of 
$\langle L / E_\nu \rangle \ll 1$ km/GeV, and therefore are much
less compelling for $\Delta m^2 \sim 1$ eV$^2$ than for larger values 
of the mass difference.

     At neutrino energies well
below the $\tau$ threshold ($E_\nu < 2$ GeV), the transition \omutau\ in the 
required region $\langle L / E_\nu \rangle \sim 1$ km/GeV will be indirectly 
probed by BooNE through $\nu_\mu$ disappearance \cite{boone}. However, a 
convincing \omutau\ signal can only be demonstrated by detecting the CC 
collisions of the $\nu_\tau$ in appearance mode. For this, an appropriate 
environment is offered by the "medium-baseline" location on Mount Jura 
\cite{icarus} in the $\nu_\mu$ beam of CERN-SPS ($L = 17$ km  and  
$\langle E_\nu \rangle = 27$ GeV by flux). The CC collisions of the 
$\nu_\tau$ may be identified either topologically \cite{icarus, nomad} or by 
detecting the detached vertex of the produced $\tau$ in a hybrid-emulsion 
spectrometer \cite{e531}. The latter option is considered in this paper. For 
the "original" transition \omue\ to be observed and investigated in the 
same experiment, that will have very different systematics compared to either
LSND \cite{lsnd} and BooNE \cite{boone}, secondary electrons must be 
efficiently detected and momentum-analyzed.

     The atmospheric evidence for neutrino oscillations driven by a mass
difference of 10$^{-2}$--10$^{-3}$ eV$^2$ \cite{atmospheric} will be
initially probed by the long-baseline experiments operating in the neutrino 
beams of proton accelerators \cite{k2k, minos, icarus}. Large angular 
divergences of these beams dictate that, for the sake of statistics, fine 
instrumentation of the detector and/or the magnetic analysis be compromised 
for a multi-kiloton fiducial mass. Therefore, unraveling the pattern of 
oscillations in this mass region will probably require a much more intense 
beam generated by a muon storage ring with a straight section that points 
towards the detector \cite{accmu}. Assuming a $\mu^-$ ring, the beam fully
consists of muon neutrinos and electron antineutrinos whose energy spectra 
and angular spreads are precisely known. That the original beam contains 
neither $\bar{\nu}_\mu$ nor $\nu_e$ allows to identify the parent neutrino 
by the sign of the lepton produced by the oscillated neutrino: negative and 
positive leptons are unambiguous signatures of the $\nu_\mu$ and 
$\bar{\nu}_e$ parents, respectively. Therefore, the transitions \omue, 
\omutau, \oemubar, and \oetaubar\ can be tagged independently.

     The unique properties of this neutrino beam may warrant a finely
instrumented detector of relatively small mass, that would be sensitive to 
a broad range of neutrino transitions driven by a mass difference of 
10$^{-2}$--10$^{-3}$ eV$^2$. For this, several conditions are obligatory.
Selecting the $\nu_\tau$ and $\bar{\nu}_\tau$ collisions by the detached 
vertex of the $\tau$ is only possible in a hybrid-emulsion apparatus. 
In order to suppress the background to $\tau$ decays arising from anticharm
production in CC collisions of the $\bar{\nu}_e$, the detector should 
identify the electrons as reliably as the muons. For the transitions \omutau\
and \oetaubar\ to be reliably discriminated, all charged secondaries (and not
just muons) should be sign-selected and momentum-analyzed in the detector.
The latter will also allow extra kinematic handles for further suppressing 
the background to $\tau$ decays (such as \ptra\ of a decay particle with
respect to $\tau$ direction and transverse disbalance of the event as a whole
with respect to incident neutrino).

     In this paper, we propose a conceptual scheme of a hybrid-emulsion
spectrometer for detecting and identifying the neutrinos of different
flavors by their CC collisions, as required for probing various channels of
neutrino oscillations. The design emphasizes detection of $\tau$ leptons by 
detached vertices, reliable identification and sign-selection of electrons, 
good spectrometry for all charged secondaries, and reconstruction of 
secondary photons and $\pi^0$ mesons. We also estimate the performance of
the proposed apparatus in a medium-baseline experiment using the neutrino
beam of the proton machine CERN-SPS and in a long-baseline experiment in the
neutrino beam of a muon storage ring.

\section{The detector}

     Building on the ideas put forth by A. Ereditato, K. Niwa, and 
P. Strolin \cite{opera}, we implement the principle of "emulsion cloud
chamber": layers of thin emulsion are only used as a tracker for events 
occurring in passive material. But in contrast with \cite{opera}, we aim at 
constructing a distributed target with low density and large radiation 
length \cite{nomad}, so that either muons, hadrons, and electrons can be 
momentum-analyzed by curvature inside the target itself. Accordingly, the 
target should largely consist of low-Z material like aluminum or carbon in 
the form of carbon-fiber composite. Apart from narrow gaps instrumented with
drift chambers that provide an electronic "blueprint" of the event, the 
target is built as a compact homogeneous volume in ambient magnetic field. 
Owing to relatively weak multiple scattering in low-Z material, the 
successive layers of thin emulsion may act as an "emulsion spectrometer" for
analyzing the momenta of charged secondaries by curvature in magnetic field. 
In untangling the topologies of neutrino events, the detector will operate 
very much like a bubble chamber. The proposed apparatus is therefore referred
to as the Emulsion Bubble Chamber, or EBC.

     Specified below is a tentative structure of the distributed target that
has been assumed in our simulations of detector response. (The actual design,
including the choice of low-Z material, must of course be carefully optimized
for a particular experiment). The fine structure of the target is depicted in 
Fig. \ref{structure}. The 6-mm-thick basic element of the structure is formed 
by a 1-mm plate of passive material, by an emulsion--plastic--emulsion sheet 
(referred to as the ES) with total thickness of 0.2 mm, and by a relatively
large drift space of 4.8 mm in which $\tau$ decays are selected. In our 
tentative design, the passive plate is largely carbon (960 $\mu$m), but also
includes a thin layer of denser substance (40 $\mu$m of copper) in order to
boost the geometric acceptance. The ES is a 100-$\mu$m sheet of transparent
plastic coated on both sides with 50-$\mu$m layers of emulsion. The medium
formed by successive elements has mean density of  $\rho = 0.49$ g/cm$^3$ and 
effective radiation length of $X_0 = 52.6$ cm. (These values of $\rho$ and 
$X_0$ are very similar to those of a bubble chamber with neon-hydrogen 
filling \cite{neon}.) Note that the element does not feature the second ES 
downstream of the gap, as originally foreseen in \cite{opera}: the idea is to 
detect the kink or the trident using track segments in the ESs of two 
successive elements, as allowed by relatively weak Coulomb scattering in 
low-Z material of the intervening passive plate.

     On the technical side, the air gap can be created by thin and rigid 
"bristles" on the upstream face of the carbon-composite plate that has been 
manufactured in "brush-like" form. (This is only possible because we have 
just one ES per element.) The positions of the thin bristles will be
tabulated, and secondary vertices that match these positions will be dropped.
Another technical option for the drift space is 5-mm-thick paper honeycomb.

     For a $\tau$ that emerged from the passive plate and suffered a 
one-prong or a three-prong decay in the gap, the decay vertex can be 
reconstructed from the track segments in two successive ESs. As soon as the 
fitted secondary vertex lies within the intervening passive plate,
the candidate event must be dropped because of the high background from 
reinteractions. In principle, a $\tau$ that decayed before reaching the
gap can be detected by impact parameter, but this possibility is not
considered here. By adding a thin layer of denser material (copper) 
downstream of low-Z material (carbon), we slightly compromise the radiation 
length for geometric acceptance: thereby, the fraction of $\tau$ leptons that
reach the air gap is effectively increased. That the proportion of copper 
events is boosted by the geometric effect is illustrated by 
Fig. \ref{vertices}. Here and below, the spectrum of incident $\tau$ 
neutrinos is assumed to be proportional to the spectrum of muon neutrinos 
from CERN-SPS \cite{nomad}.

     For the tracks to be found in emulsion at scanning stage, planes of
electronic detectors must be inserted in the continuous structure formed
by the basic elements. These detectors should allow a crude on-line 
reconstruction of the event as a whole, and therefore they must have
sufficient spatial resolution, provide angular information, and be spaced
by less than one radiation length $X_0$. In our tentative design, a stack
of 30 elements forms a "module" with total thickness of $0.34 X_0$, and
4-cm-wide gaps between adjacent modules are instrumented by multisampling
drift chambers.

     In a medium- or long-baseline experiment where the occupancy of ESs 
will be relatively low, the accuracy of "electronic" reconstruction should be
sufficient for unambiguously finding an energetic track in emulsion. 
Therefore, one can scan back along a stiff track, starting from the 
downstream ES of the module in which the collision occurred. As soon as the 
layer of origin is reached, a few successive ESs of the nearby elements must 
be fully scanned over relatively small areas towards finding the stubs of 
all other tracks associated with the primary vertex. To refine the alignment
of ESs, a sufficient number of stiff muon tracks must be fully reconstructed
in emulsion. This first stage of emulsion scanning will yield a relatively 
small number of events featuring decay signatures (either a kink or a trident
in the air gap just downstream of the primary vertex). The second stage will 
be to scan down all tracks emerging from the primary vertex and to find and 
measure the conversions of secondary photons in the distributed target. This
will allow to analyze the momenta of the secondaries in the "emulsion
spectrometer" and to identify electrons by change of curvature and by 
emission of brems.

 We foresee that the scanning of emulsion will only start 
after the full period of detector exposure.

\section{Spectrometry for charged particles and photons}

     The response of the EBC detector is simulated using GEANT. In fitting a 
track, we assume that each tracker traversed (either a ES or a drift chamber)
provides two spatial points with resolutions of 2 and 150 $\mu$m, 
respectively. Because of multiple scattering, the fit is but marginally 
sensitive to increasing the spatial error in the ES from 2 up to 8 $\mu$m.
A uniform magnetic field of 0.7 Tesla, that is normal with respect to beam
direction, is assumed throughout. (Using a stronger field would boost the
performance of the spectrometer, but may be impracticable for a big air-core
magnet that will house a multiton EBC.)

     When found in emulsion, the track of an electron can be identified by
a variation of curvature due to energy losses in successive layers of the
target. These losses must be accounted for in estimating the momentum by
curvature in magnetic field. For this, we use a simple algorithm that should 
be treated as preliminary and is not fully realistic. Namely, we fit
a restricted segment of the $e^-$ track over which the actual loss of energy 
does not exceed 30\%. 

Additionally, this segment is required to cross no less than five ESs.
We compute an "ideal trajectory" for a given value of electron momentum  
$p_e$, to which the observed trajectory is then fitted, using GEANT. In doing
so, we switch off multiple scattering and radiation losses, but instead 
reduce the momentum "by hand" in each layer of the target by the same amount 
$\Delta p$  that is treated as an empirical parameter. The value of  
$\Delta p$ is selected so as to obtain (i) an unbiased estimate of electron 
momentum ($\langle p_e^{\mathrm{meas}} / p_e^{\mathrm{true}} \rangle \sim 1$)
and (ii) a reasonable value of $\chi^2$. 

     For definiteness, we use a Monte-Carlo template of electrons with
$p_e > 1$ GeV originating from simulated decays \xedec\ (see further). For 
these electrons, the mean length of the selected track segment is close to 
30 cm (see Fig. \ref{length}), which proves to be sufficient for analyzing 
the momentum in the "emulsion spectrometer" formed by successive ESs of the 
distributed target. Plotted in Fig. \ref{aluminum} is the ratio between the 
fitted and true momenta, $p_e^{\mathrm{meas}} /p_e^{\mathrm{true}}$, for 
the properly fitted electrons ($\chi^2 < 3$). The same ratio is then 
separately shown for two regions of electron momentum: $1 < p_e < 5$ GeV and 
$p_e > 5$ GeV. We estimate that of all electrons with $p_e > 1$ GeV, nearly 
90\% can be reliably detected, identified, and sign-selected (that is, 
reconstructed with  $\chi^2 < 3$  and $\delta p_e / p_e < 0.40$). On average,
the $e^-$ momentum is measured to a precision of some 11\%. 

     Treating the length of the track segment as a free parameter of the fit,
which is a more realistic approach that is not attempted in this paper, will 
further improve the momentum resolution and boost the fraction of 
sign-selected electrons. Good spectrometry for electrons will allow EBC to 
detect and discriminate the CC collisions of electron neutrinos and 
antineutrinos and, at the same time, to select electronic decays of the 
$\tau$ almost as efficiently as the muonic ones. A comparison between the two
leptonic modes will provide an important handle on self-consistency of any 
$\tau$ signal.

     The spectrometry of muons and charged pions is much less affected by 
energy loss in matter, but for pions the momentum resolution is slightly 
degraded by hadronic reinteractions. For the pions with $p_\pi > 1$ GeV 
originating from the simulated decays \xpidec, Fig. \ref{momres} shows the 
ratio between the fitted and true momenta, 
$p_\pi^{\mathrm{meas}} /p_\pi^{\mathrm{true}}$. The mean uncertainty on pion 
momentum is seen to be close to 7\%.

     Neutral pions can be reconstructed from photon conversions in the 
distributed target. (Here, we assume that the potential length in the 
detector is much larger than $X_0 = 52.6$ cm.) In estimating the energy of a 
photon that has converted in the target, we count only those conversion 
electrons that have fired at least one drift chamber and, therefore, can be 
found in emulsion and then momentum-analyzed by curvature. We also assume 
that the primary vertex has already been fitted, so that the direction of the
photon is precisely known. For illustration, we consider the $\pi^0$ mesons
originating from the decay \zrhodec. The measured invariant mass of the two 
detected photons from \ypi_gg, as plotted in Fig. \ref{mgg}, shows a distinct 
$\pi^0$ signal. The actual size of the mass window for selecting the $\pi^0$ 
candidates will be dictated by the level of combinatorial background in a 
particular analysis; for purely illustrative purposes, we assume a mass 
window of $115 < m_{\gamma\gamma} < 155$ MeV. Thus estimated detection 
efficiency for $\pi^0$ mesons is close to 0.26.

\section{Detecting the leptonic and semileptonic decays of the $\tau$}

     For the $\tau$ leptons emitted in either the deep-inelastic and 
quasielastic $\nu_\tau N$ collisions, we generate the two leptonic decays,
\xmudec\ and \xedec, and three semileptonic (quasi-)two-body decays: \xpidec,
\zrhodec\ that is mediated by the resonance \yrho_2pi, and \za1dec\ that is 
mediated by the resonance \ya1_3pi. The threshold effect for $\tau$ 
production in neutrino--nucleon collisions and polarization of the $\tau$, 
that affects the angular distribution of decay products in the $\tau$ frame, 
are accounted for. Further details on our $\tau$ generator can be found 
in \cite{ourselves}.

     For definiteness, we again assume that $\tau$ neutrinos incident on EBC 
have the same energy spectrum as muon neutrinos from CERN-SPS \cite{nomad}
($\langle E_\nu \rangle = 27$ GeV by flux),
and that the distributed target is magnetized by a uniform field of 0.7 Tesla
that is perpendicular to beam direction. The decay products of the $\tau$ are
propagated through the target and then reconstructed from hits in emulsion,
as explained in the previous section. In estimating the detection 
efficiencies for different decay channels of the $\tau$, we adopt the 
(quasi-realistic) selection criteria that are listed below.
\begin{itemize}
\item

The $\tau$ must have decayed in the drift gap. This is necessary for
reconstructing the detached vertex from track segments in the upstream and
downstream ESs.

\item

The momentum of either charged daughter must exceed 1 GeV, and its emission 
angle must lie within 400 mrad of beam direction. These selections are 
suggested by the fact that soft and broad-angle tracks are poorly 
reconstructed in emulsion.

\item

For a one-prong decay to a charged daughter $d$ (either a muon, electron, or 
pion), the kink angle must be suffiently large: $\theta_{\tau d} > 20$ mrad.
This lower cut reflects the experimental uncertainty on the kink angle
that is close to 5 mrad.

\item

All charged daughters of the $\tau$ must be momentum-analyzed by curvature
and reliably sign-selected (that is, reconstructed in the detector with  
$\chi^2 < 3$  and  $\delta p / p < 0.40$). This will fix the sign of the
$\tau$ and, on the other hand, will allow kinematic handles for rejecting 
the decays of charmed and strange particles.

\item

In a one-prong decay, \ptra\ of the charged daughter with respect to $\tau$ 
direction must exceed 250 MeV. This is aimed at rejecting the decays of
strange particles.

\end{itemize}

     For those $\tau$ decay channels that have actually been simulated in the 
detector, the estimated detection efficiencies (or acceptances) are listed in
Table \ref{accept}. (Note that here we do not require the $\pi^0$ from 
\zrhodec\ to be reconstructed.) For these decay channels of the $\tau$, the 
acceptance-weighted branching fractions add up to some 0.28. Approximately
accounting for the other one-prong and three-prong channels, we estimate that
nearly 32\% of all $\tau$ leptons produced in passive material will be 
detected in EBC by visible kinks or tridents.

\begin{table}[p]
\begin{tabular}{|c|c|c|c|c|}
\hline
Decay channel & Branching & Acceptance           & Acceptance \\
              & fraction  &                      & times     \\
              &           &                      & branching  \\
\hline
   \xmudec\   &  0.174    &    0.40             &  0.070    \\
\hline
   \xedec\    &  0.178    &    0.36             &  0.064   \\
\hline
   \xpidec\   &  0.113    &    0.40             &  0.045   \\
\hline
   \zrhodec\  &  0.252    &    0.30             &  0.076   \\
\hline
   \za1dec\   &  0.094    &    0.27             &  0.025   \\
\hline
\end{tabular}
\caption{The detection efficiencies, or acceptances, for those decay 
channels of the $\tau$ that have been simulated in the detector.}
\label{accept}
\end{table}

     The (quasi-)two-body decays \xpidec, \xrhodec, and \xa1dec\ may
provide an extra kinematic handle for discriminating the $\tau^-$ against
the background of anticharm. In a decay  $\tau^- \rightarrow h^-\nu$, 
"transverse mass" is defined as 
$M_{\mathrm{T}} = \sqrt {m_h^2 + p_{\mathrm{T}}^2} + p_{\mathrm{T}}$, 
where $m_h$ and \ptra\ are the $h^-$ mass and transverse momentum with 
respect to $\tau$ direction. The two-body kinematics dictate that the 
unsmeared \mtra\ distribution should reveal a very distinctive peak just 
below  $M_{\mathrm{T}}^{\mathrm{max}} = m_\tau$, see the upper plots of
Fig. \ref{m_t}. The \mtra\ technique for identifying massive parents by 
two-body decays in emulsion was proven by discriminating the relatively 
rare decay $D_s^+(1968) \rightarrow \mu^+ \nu$  against a heavy background 
from other decays of charm \cite{e653}. Observing the high-\mtra\ peak in 
a detector requires good spectrometry of charged pions and, for the 
decay \zrhodec\ in particular, good reconstruction of $\pi^0$ mesons.
The bottom plots of Fig. \ref{m_t} show the smeared \mtra\ distributions
for the reconstructed decays \xpidec, \zrhodec, and \za1dec\ (the $\pi^0$
for \zrhodec\ has been selected in the mass window
$115 < m_{\gamma\gamma} < 155$ MeV, see Fig. \ref{mgg}). The Jacobian 
cusps of the original \mtra\ distributions largely survive the apparatus 
smearings of EBC and, therefore, will provide distinctive signatures of 
the $\tau$.

\section{Sensitivity to neutrino oscillations}

     Neutrino oscillations driven by $\Delta m^2 \sim 1$ eV \cite{lsnd} can be 
efficiently probed at a location on Mount Jura near CERN, that is irradiated 
by the existing wide-band $\nu_\mu$ beam of the CERN-SPS accelerator 
($\langle E_\nu \rangle = 27$ GeV by flux) over a "medium" baseline of 17 km 
\cite{icarus}. At this location, an EBC detector can be deployed in the 
existing magnet of the NOMAD detector \cite{nomad} that delivers a magnetic
field of up to 0.7 Tesla. In the large internal volume of this magnet, 
$3.5\times3.5\times7.5$ m$^3$, one might deploy a distributed target with a 
volume of $3.0 \times 3.0 \times 6.6$ m$^3$, that will consist of 30 modules 
(see Section 2). The target has a total thickness of 10 radiation lengths, 
and contains nearly 20 tons of passive material and 3 tons of standard 
emulsion (or a lesser amount of diluted emulsion, that may be warranted by a 
relatively low occupancy at a medium-baseline location). The magnetic volume
will also house a few drift chambers downstream of the target. As photon 
conversions and electrons will be efficiently detected in the distributed
target, no extra electromagnetic calorimeter is foreseen. The design 
of the muon system is not discussed in this paper.

     At the Jura location, the rate of $\nu_\mu$-induced CC collisions has 
been estimated \cite{icarus} as 843 events per ton of target per $10^{19}$ 
protons delivered by CERN-SPS. Assuming $10^{20}$ delivered protons which 
corresponds to 3--4 years of operation, the proposed detector will collect 
nearly $1.7\times10^5$ CC events. Even if the probability of the \omutau\ 
transition is as small as 0.3\% \cite{lsnd}, we will detect some 80 $\tau$ 
events with a negligibly small background from the decays of strange and 
anticharm particles. (This prediction assumes a $\tau$ detection efficiency
of 32\%, as estimated above for the same beam, and takes into account the
threshold effect in $\tau$ production.) Alternatively, a zero signal will 
allow to exclude (at 90\% C.L.) an area of the parameter plane
($\sin^2 2\theta_{\mu\tau}$, $\Delta m^2_{\mu\tau}$) that is depicted in 
Fig. \ref{exclude}.

     As demonstrated above, EBC will very efficiently identify, sign-select, 
and momentum-analyze prompt electrons from CC collisions of electron
neutrinos and antineutrinos. The energy of incident $\nu_e$ 
($\bar{\nu_e}$) will be estimated to better than 10\%. If at the Jura site 
the transition \omue\ indeed occurs with a probability of some 0.003 as in 
LSND \cite{lsnd}, in an exposure of $10^{20}$ protons on target we will 
detect and reconstruct some 460 $\nu_e N \rightarrow e^-X$ events due to 
oscillated neutrinos against a background of nearly 860 CC events due to the
original $\nu_e$ component of the beam \cite{icarus}. As the original 
$\nu_e$ component of the beam is substantially harder than the $\nu_\mu$ 
component, a \omue\ signal will effectively reduce the mean energy of  
$\nu_e N \rightarrow e^-X$ events.

     A very intense and collimated beam of a muon storage ring \cite{accmu}
will allow long-baseline experiments with relatively small but finely
instrumented detectors. For purely illustrative purposes, we assume an EBC
detector as sketched above (20 tons of passive material plus 3 tons of
standard emulsion) that is irradiated by a $\mu^-$ ring with parameters 
as foreseen in \cite{accmu} (that is, $7.5 \times 10^{20}$ injected muons per
year of operation and a straight section that amounts to 25\% of the ring 
circumference). We also assume a "nominal" distance of 732 km (either from 
CERN to Gran Sasso or from Fermilab to Soudan) that results in effective 
baselines of $\langle L / E_\nu \rangle \sim 20$ and 10 km/GeV for the
stored-muon energies of $E_\mu = 50$ and 100 GeV, respectively. Then, given
unpolarized muons with $E_\mu = 50$ (100) GeV in the ring, in the absence of 
oscillations the detector will annually collect some  $5.5 \times 10^3$
($4.4 \times 10^4$) and $2.4 \times 10^3$ ($1.9 \times 10^4$)  CC collisions 
of muon neutrinos and electron antineutrinos, respectively. Note that the
expected event rate for $E_\mu = 100$ GeV is as high as in the CERN-SPS beam 
at Jura location.

     Taken together, the data on atmospheric \cite{atmospheric} and reactor 
\cite{chooz} neutrinos favor a \omutau\ transition with almost maximal mixing
and with a mass difference in the range 
$\Delta m^2_{\mu\tau} \sim 10^{-2}$--10$^{-3}$ eV$^2$ 
\cite{gonzalez, yasuda, fogli}.
For definiteness assuming $\sin^2 2\theta_{\mu\tau} = 1$  and  
$\Delta m^2_{\mu\tau} = 5 \times 10^{-3}$ eV$^2$, we estimate that in 
three years of operation a 20-ton EBC will detect 
a signal of some 50 (170) $\tau^-$ leptons for $E_\mu = 50$ (100) GeV with a 
negligible background of $\bar{\nu}_e$-produced anticharm. (Note that the
$E_\mu$-dependence of the $\tau$ signal largely reflects the increase of
$\sigma(\nu_\tau N \rightarrow \tau^- X)$ with neutrino energy: in the region
$\Delta m^2$(eV$^2$) $\ll \langle E_\nu \rangle / L$ (GeV/km), the flux of
oscillated neutrinos is virtually independent of $E_\mu$.) Alternatively, a 
zero $\tau^-$ signal will effectively exclude a large area of parameter space 
for the transition \omutau, see Fig. \ref{exclude}. The transition \oetaubar\ 
will be simultaneously detected with a comparable sensitivity, and the two 
transitions will be reliably discriminated by the charge of the $\tau$. Using
the detached-vertex information, the transitions \omue\ and \oemubar\ will be 
separated from the transitions \omutau\ and \oetaubar\ involving the leptonic
decays \xedec\ and $\tau^+ \rightarrow \mu^+\nu\bar{\nu}$, respectively. We 
may conclude that, in this experimental environment, even a relatively small 
EBC detector will efficiently probe the entire pattern of neutrino 
oscillations in the region $\Delta m^2 \sim 10^{-2}$--10$^{-3}$ eV$^2$.

\section{Summary}

     A conceptual detector scheme is proposed for studying various channels
of neutrino oscillations. The hybrid-emulsion spectrometer will detect and 
discriminate the neutrinos of different flavors by their CC collisions. The 
design emphasizes detection of $\tau$ leptons by detached vertices, 
identification and sign-selection of electrons, and spectrometry for all 
charged particles and photons. A distributed target is formed by layers of 
low-Z material, emulsion-plastic-emulsion sheets, and air gaps in which 
$\tau$ decays are detected. Target modules with mean density of 0.49 g/cm$^3$
and radiation length of 52.6 cm, that are similar to those of a bubble 
chamber with neon-hydrogen filling, are alternated by multisampling drift 
chambers that provide electronic tracking in real time. The tracks of charged
secondaries, including electrons, are momentum-analyzed by curvature in 
magnetic field using hits in successive thin layers of emulsion and in drift 
chambers. Electrons are identified by change of curvature and by emission of 
brems. Photons are detected and analyzed by conversions in the distribute
target, and neutral pions are reconstructed. The $\tau$ leptons are 
efficiently detected and sign-selected in all major decay channels, 
including \xedec.

     At a medium-baseline location on mount Jura in the existing neutrino 
beam of the proton machine CERN-SPS, the detector will be sensitive to either
the \omutau\ and \omue\ transitions in the mass-difference region of
$\Delta m^2 \sim 1$ eV$^2$, as suggested by the results of LSND.
At a long-baseline location in the neutrino beam of a muon storage ring, even 
a relatively small spectrometer of the proposed type will efficiently probe 
the entire pattern of neutrino oscillations in the region
$\Delta m^2 \sim 10^{-2}$--10$^{-3}$ eV$^2$ that is suggested by the data
on atmospheric neutrinos.

     This work was supported in part by the CRDF foundation (grant RP2-127)
and by the Russian Foundation for Fundamental Research (grant 98-02-17108).

\newpage
\noindent {\bf Figure Captions}

\begin{figure}[hbt]
\vspace{20 cm}
\includegraphics{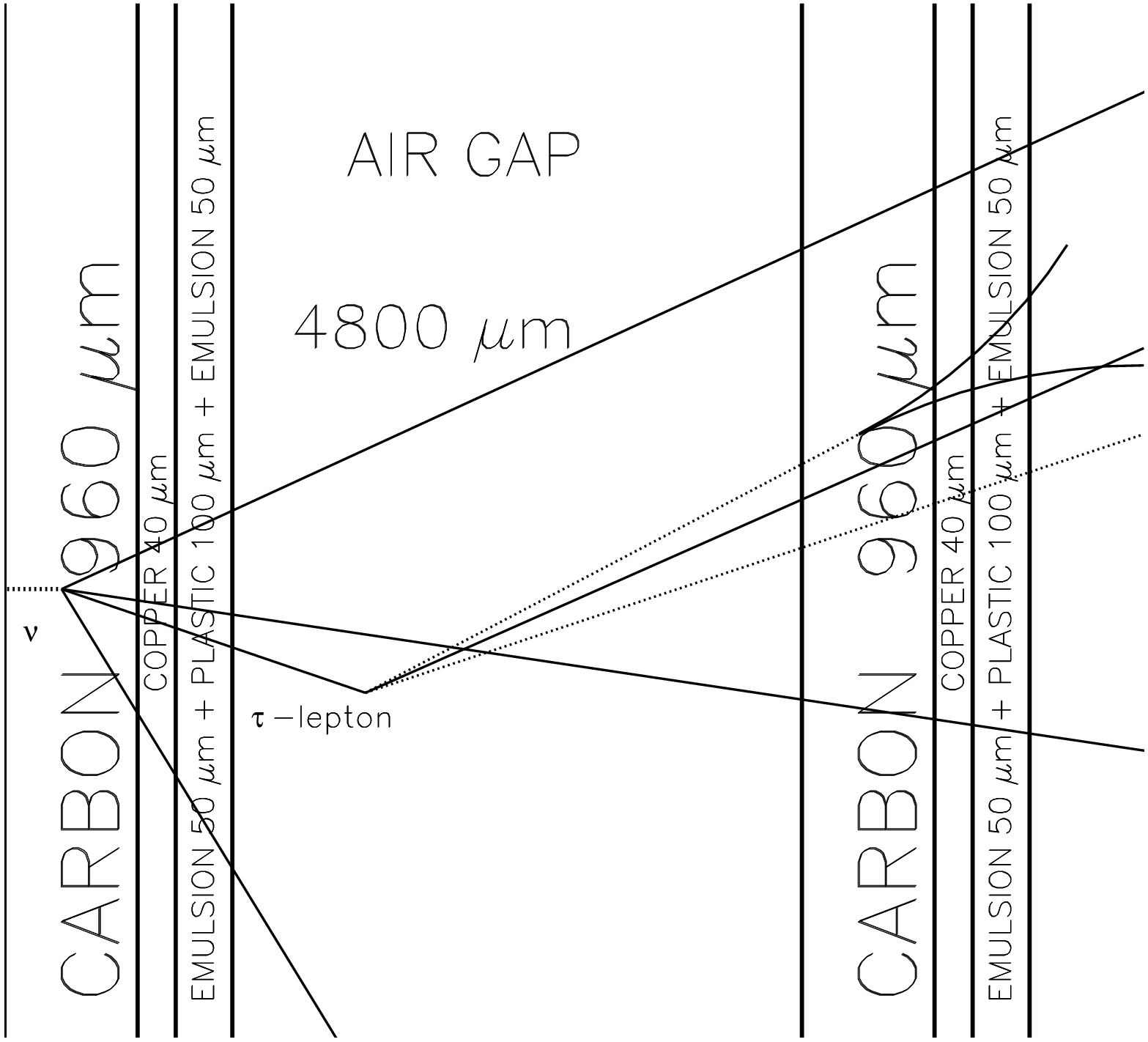} 
\caption { 
Schematic of the assumed fine structure of the target, showing the 
1-mm-thick carbon--copper plates (960 + 40 $\mu$m),
emulsion--plastic--emulsion sheets (50 + 100 + 50 $\mu$m), and drift space
in which  $\tau$ decays are selected (4800 $\mu$m).}
\label{structure}
\end{figure}

\noindent Fig. \ref{structure}. \\
Schematic of the assumed fine structure of the target, showing the 
1-mm-thick carbon--copper plates (960 + 40 $\mu$m), 
emulsion--plastic--emulsion sheets (50 + 100 + 50 $\mu$m), and drift space
in which $\tau$ decays are selected (4800 $\mu$m).

\begin{figure}[hbt]
\vspace{15 cm}
\includegraphics{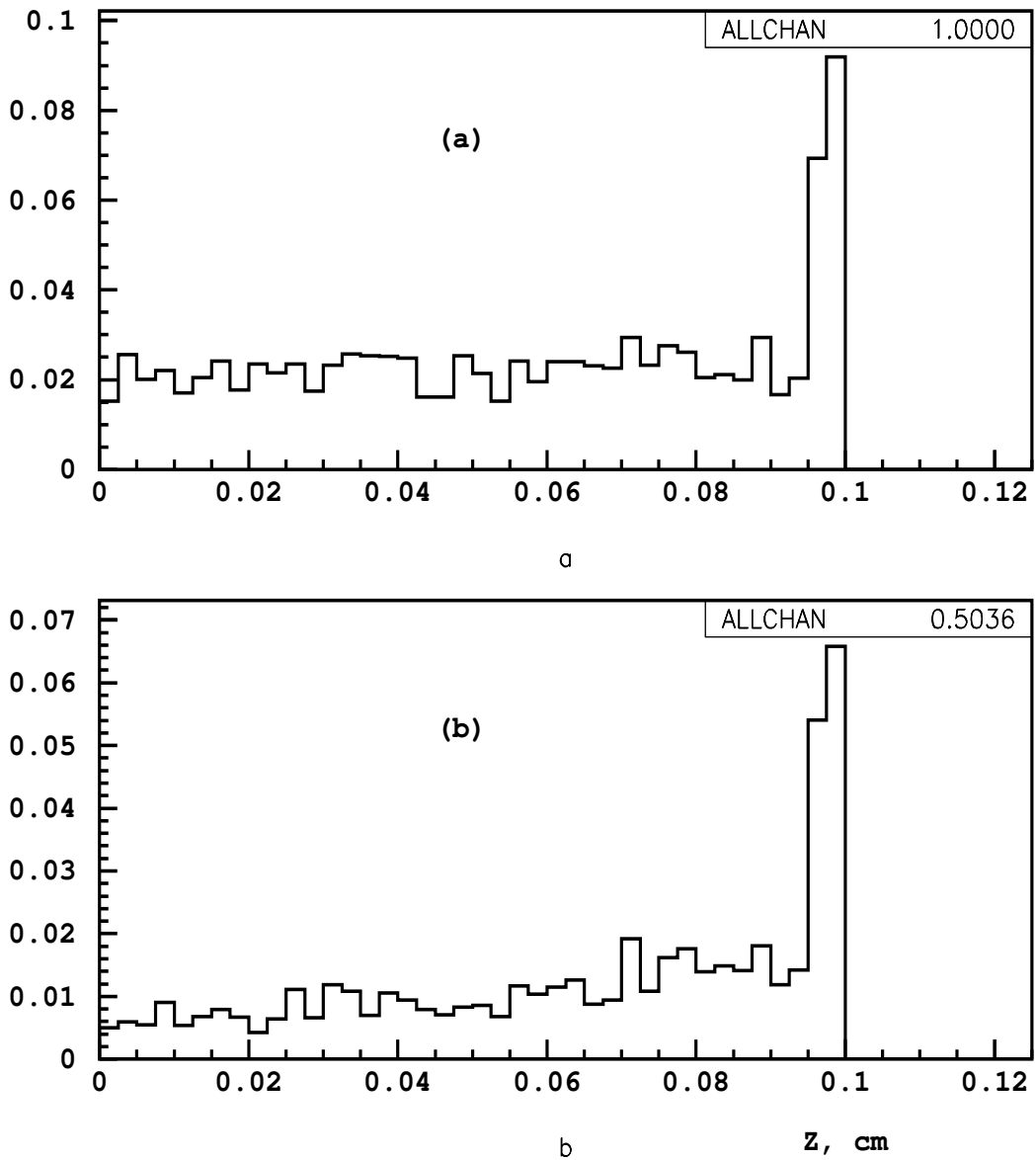}
\caption { 
The depth of the primary vertex in the carbon--copper
plate for all $\tau$ events (a) and for those events in which the $\tau$ has
decayed in the drift space (b). Here and in subsequent figures, the energy 
spectrum of incident $\tau$ neutrinos is assumed to be proportional to the
spectrum of muon neutrinos from the CERN-SPS accelerator.}
\label{vertices}
\end{figure}

\noindent Fig. \ref{vertices}. \\
The depth of the primary vertex in the carbon--copper
plate for all $\tau$ events (a) and for those events in which the $\tau$ has 
decayed in the drift space (b). Here and in subsequent figures, the energy
spectrum of incident $\tau$ neutrinos is assumed to be proportional to the
spectrum of muon neutrinos from the CERN-SPS accelerator.

\begin{figure}[hbt]
\vspace{21 cm}
\includegraphics{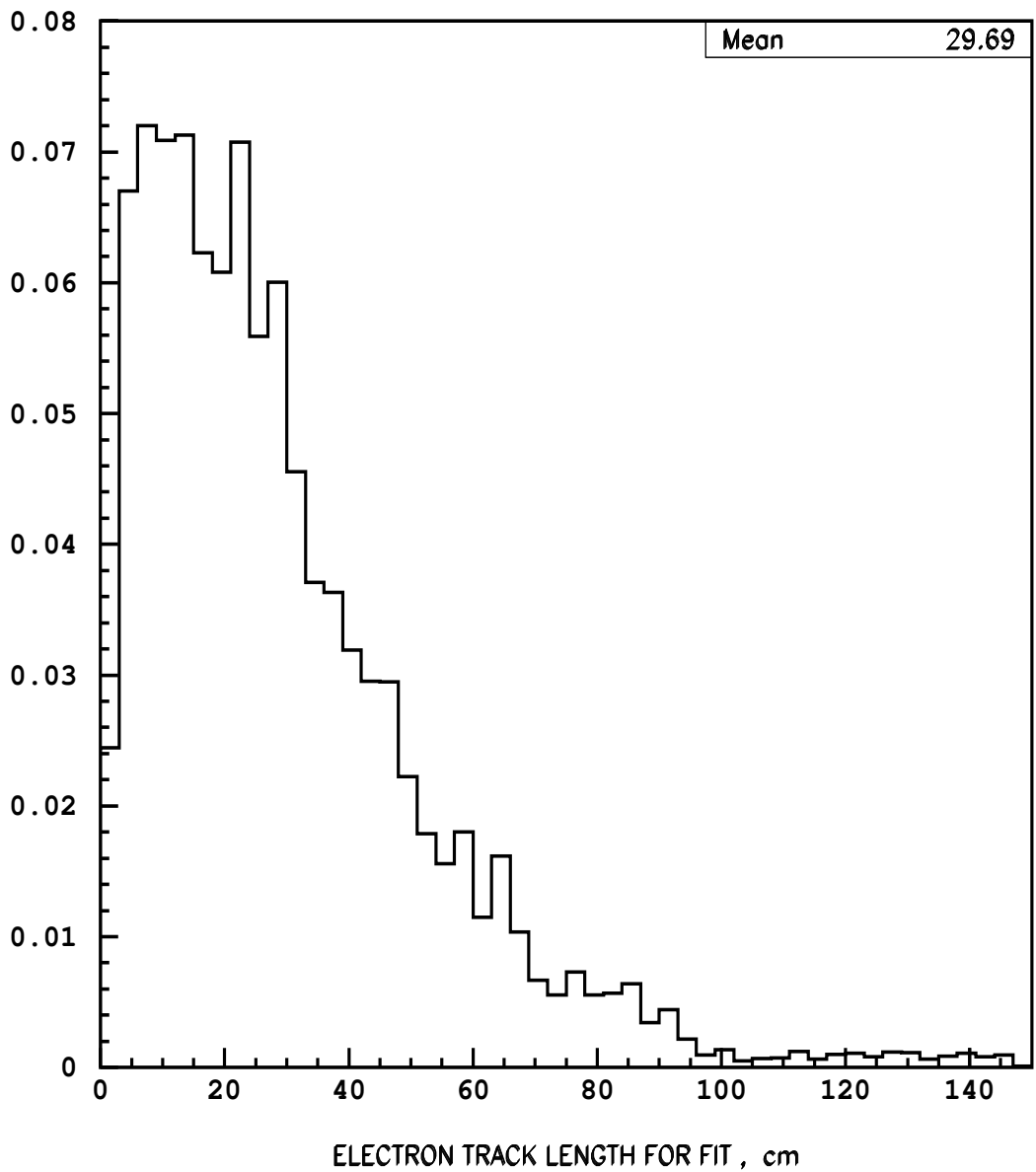} 
\caption{ 
The length along the beam direction of the track segment used for analyzing
the $e^-$ momentum by curvature (see text).}
\label{length}
\end{figure}

\noindent Fig. \ref{length} \\
The length along the beam direction of the track segment used for analyzing
the $e^-$ momentum by curvature (see text).

\begin{figure}[hbt]
\vspace{20 cm}
\includegraphics{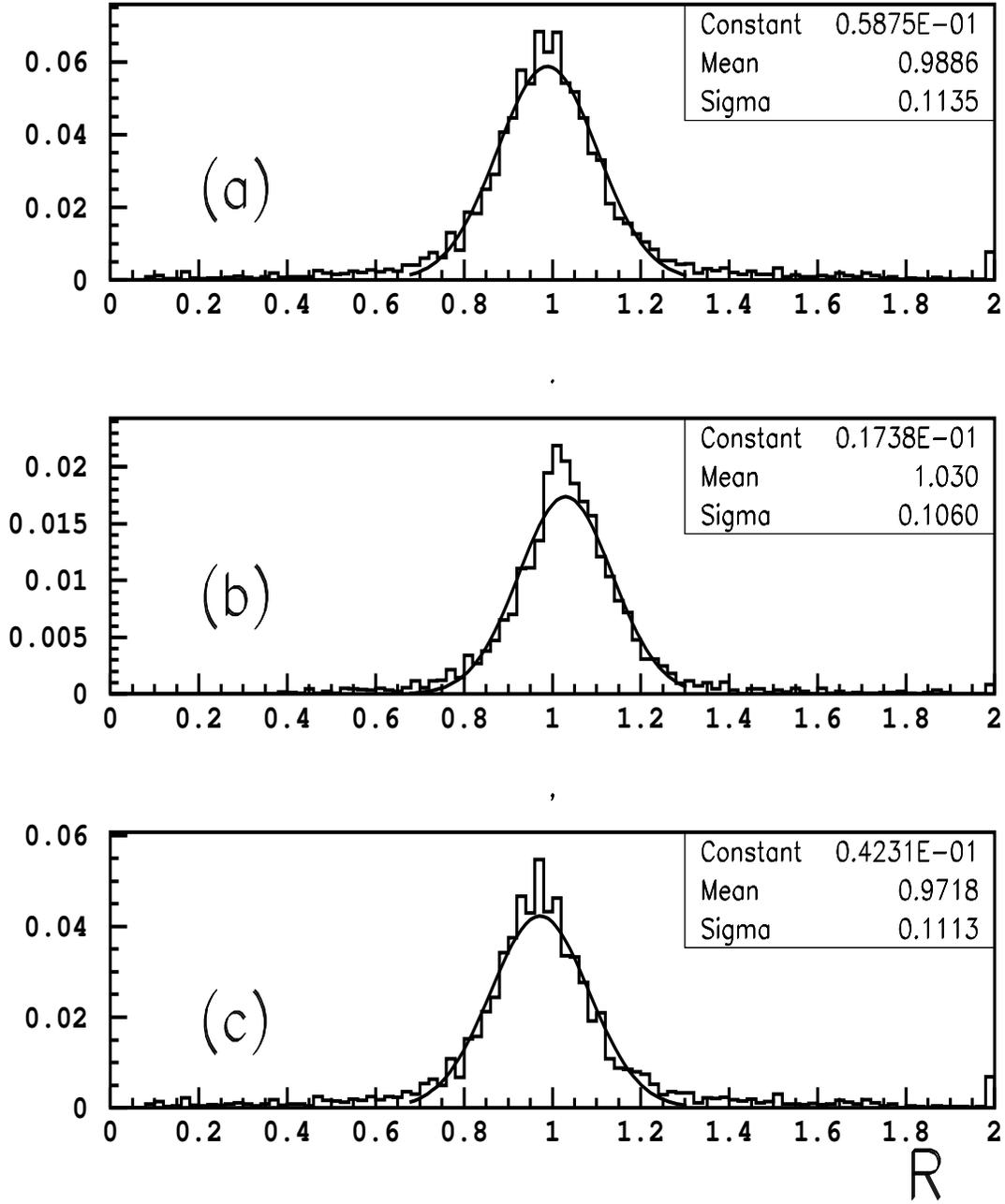} 
\caption{ 
The ratio between the fitted and true momenta of the electron from
\xedec, $R = p_e^{\mathrm{meas}} / p_e^{\mathrm{true}}$, for 
$p_e > 1$ GeV (a), for $1 < p_e < 5$ GeV (b), and for  $p_e > 5$ GeV (c).}
\label{aluminum}
\end{figure}

\noindent Fig. \ref{aluminum}. \\
The ratio between the fitted and true momenta of the electron from
\xedec, $R = p_e^{\mathrm{meas}} / p_e^{\mathrm{true}}$, for 
$p_e > 1$ GeV (a), for $1 < p_e < 5$ GeV (b), and for  $p_e > 5$ GeV (c).

\begin{figure}
\vspace{15 cm}
\includegraphics{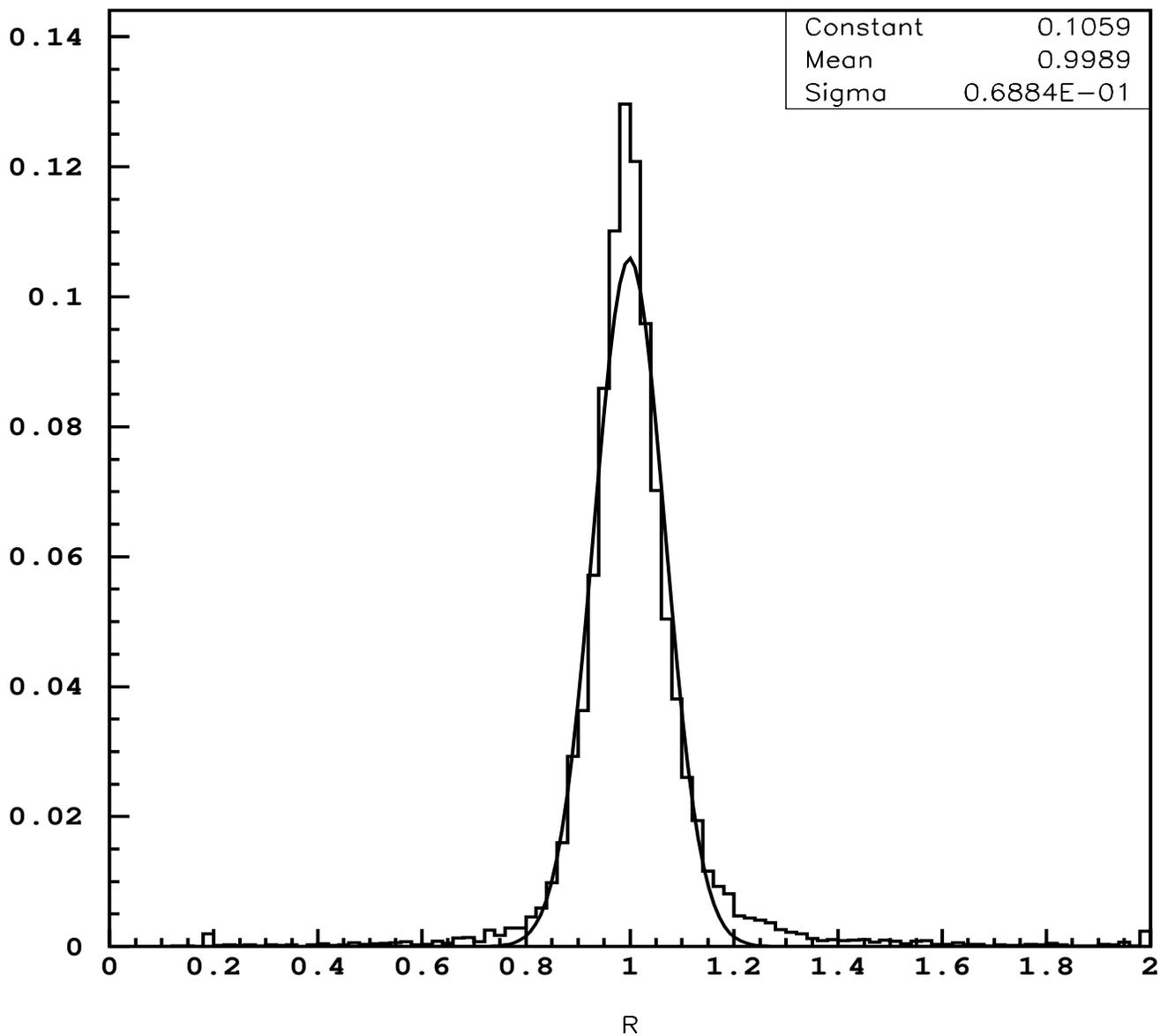} 
\caption{ 
The ratio between the fitted and true momenta,
$R = p_\pi^{\mathrm{meas}} / p_\pi^{\mathrm{true}}$, for the $\pi^-$ with
$p_\pi > 1$ GeV originating from the decay \xpidec.}
\label{momres}
\end{figure}

\noindent Fig. \ref{momres}. \\
The ratio between the fitted and true momenta,
$R = p_\pi^{\mathrm{meas}} / p_\pi^{\mathrm{true}}$, for the $\pi^-$ with
$p_\pi > 1$ GeV originating from the decay \xpidec.

\begin{figure}
\vspace{19 cm}
\includegraphics{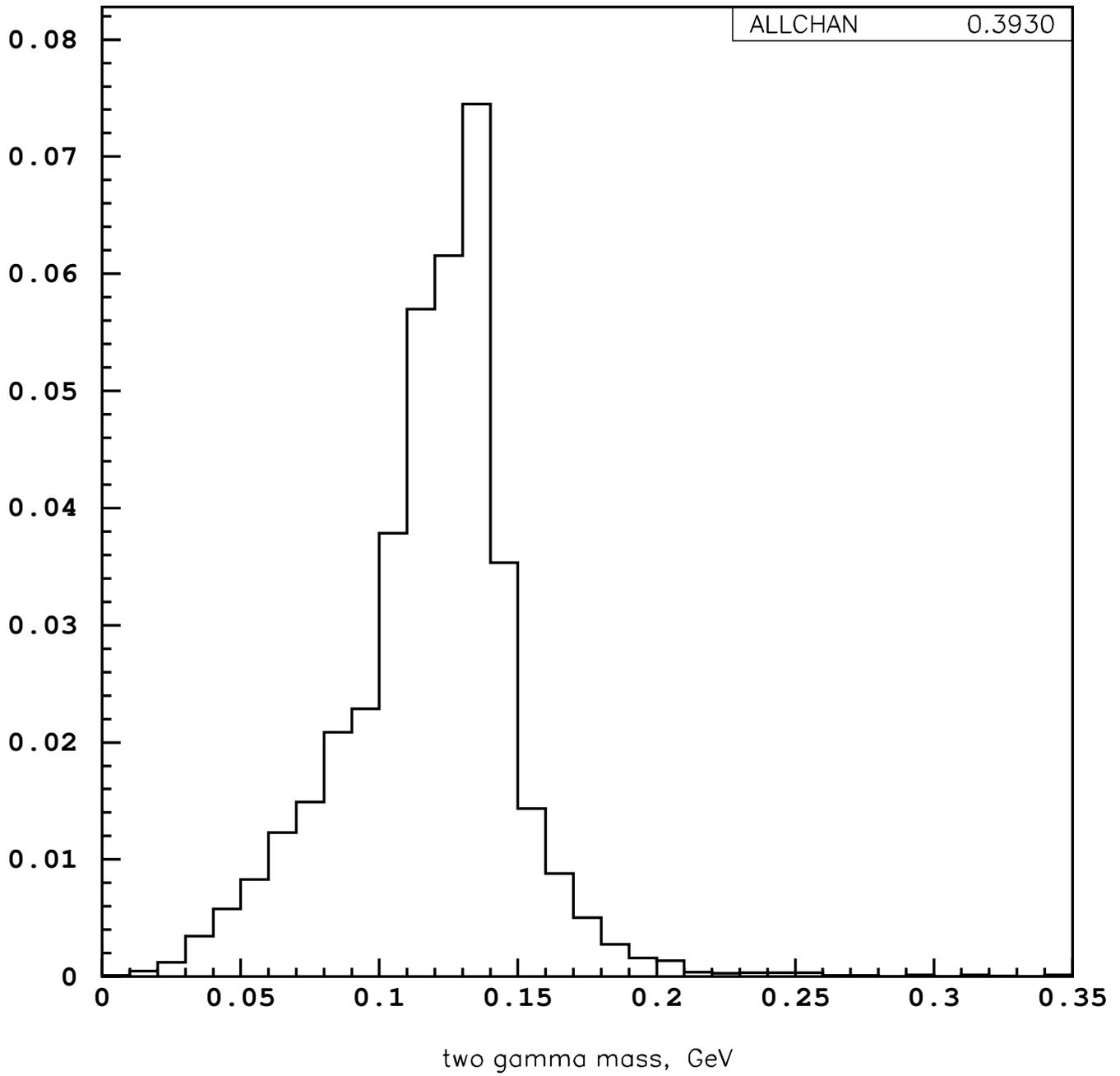} 
\caption{ 
For the decay  \zrhodec\ followed by \ypi_gg, the measured invariant
mass of the two photons that have been detected by conversions in the
distributed target.}
\label{mgg}
\end{figure}

\noindent Fig. \ref{mgg}. \\
For the decay  \zrhodec\ followed by \ypi_gg, the measured invariant
mass of the two photons that have been detected by conversions in the 
distributed target.

\begin{figure}
\vspace{19 cm}
\includegraphics{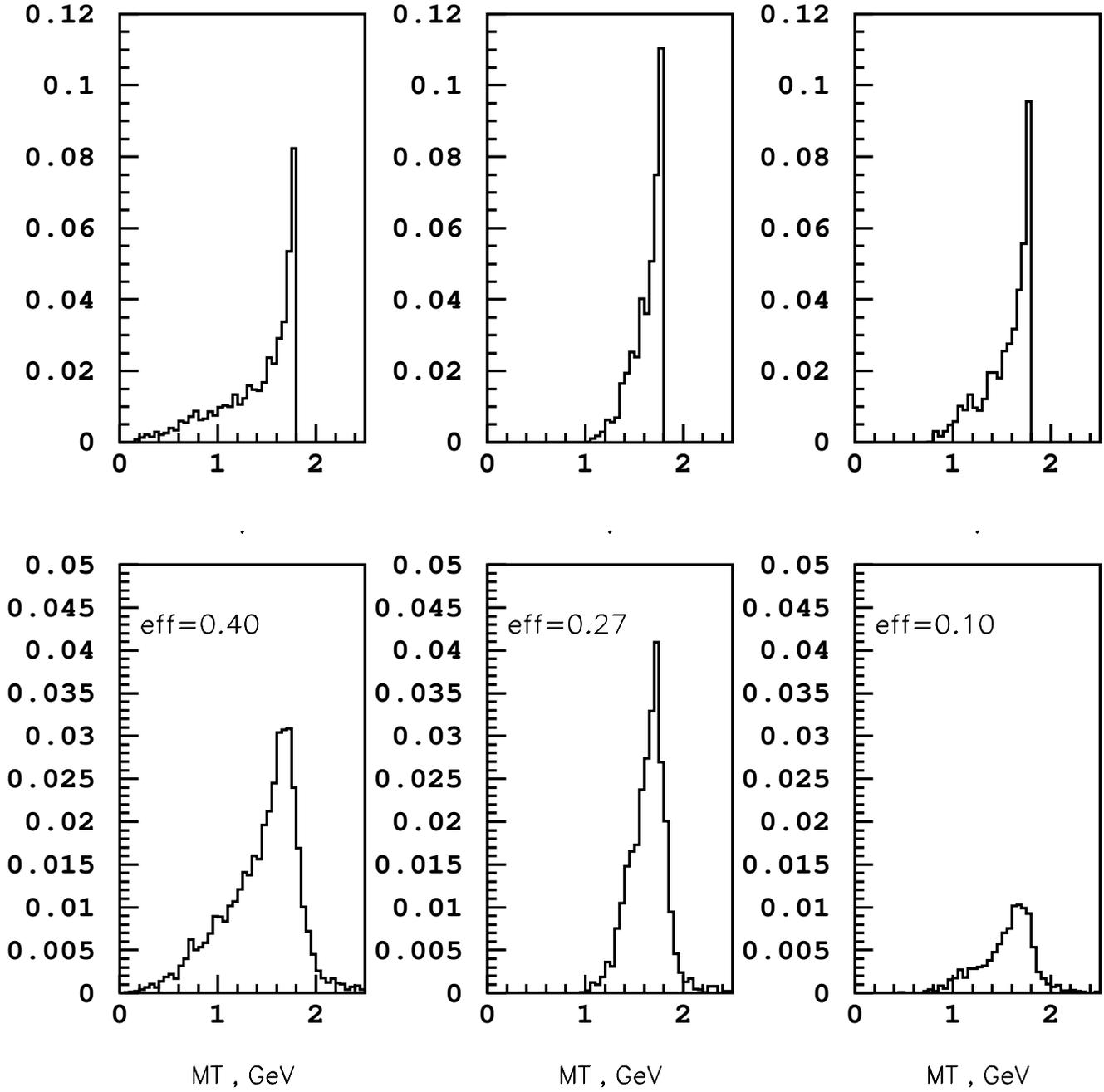} 
\caption{ 
Transverse mass
$M_{\mathrm{T}} = \sqrt {m_h^2 + p_{\mathrm{T}}^2} + p_{\mathrm{T}}$ for the
(quasi-)two-body decays $\tau^- \rightarrow h^-\nu$  with 
$h^- = \pi^-$ (left-hand column), 
$h^- = a_1^-\rightarrow\pi^-\pi^+\pi^-$ (middle column), and
$h^- = \rho^-\rightarrow\pi^-\pi^0$ (right-hand column).
The unsmeared \mtra\ distributions for all events in each channel prior to 
any selections are shown in the top row. The smeared distributions for
detected events are shown in the bottom row.}
\label{m_t}
\end{figure}

\noindent Fig. \ref{m_t}. \\
Transverse mass  
$M_{\mathrm{T}} = \sqrt {m_h^2 + p_{\mathrm{T}}^2} + p_{\mathrm{T}}$ for the
(quasi-)two-body decays $\tau^- \rightarrow h^-\nu$  with 
$h^- = \pi^-$ (left-hand column), 
$h^- = a_1^-\rightarrow\pi^-\pi^+\pi^-$ (middle column), and
$h^- = \rho^-\rightarrow\pi^-\pi^0$ (right-hand column).
The unsmeared \mtra\ distributions for all events in each channel prior to 
any selections are shown in the top row. The smeared distributions for 
detected events are shown in the bottom row.

\begin{figure}
\vspace{18 cm}
\includegraphics{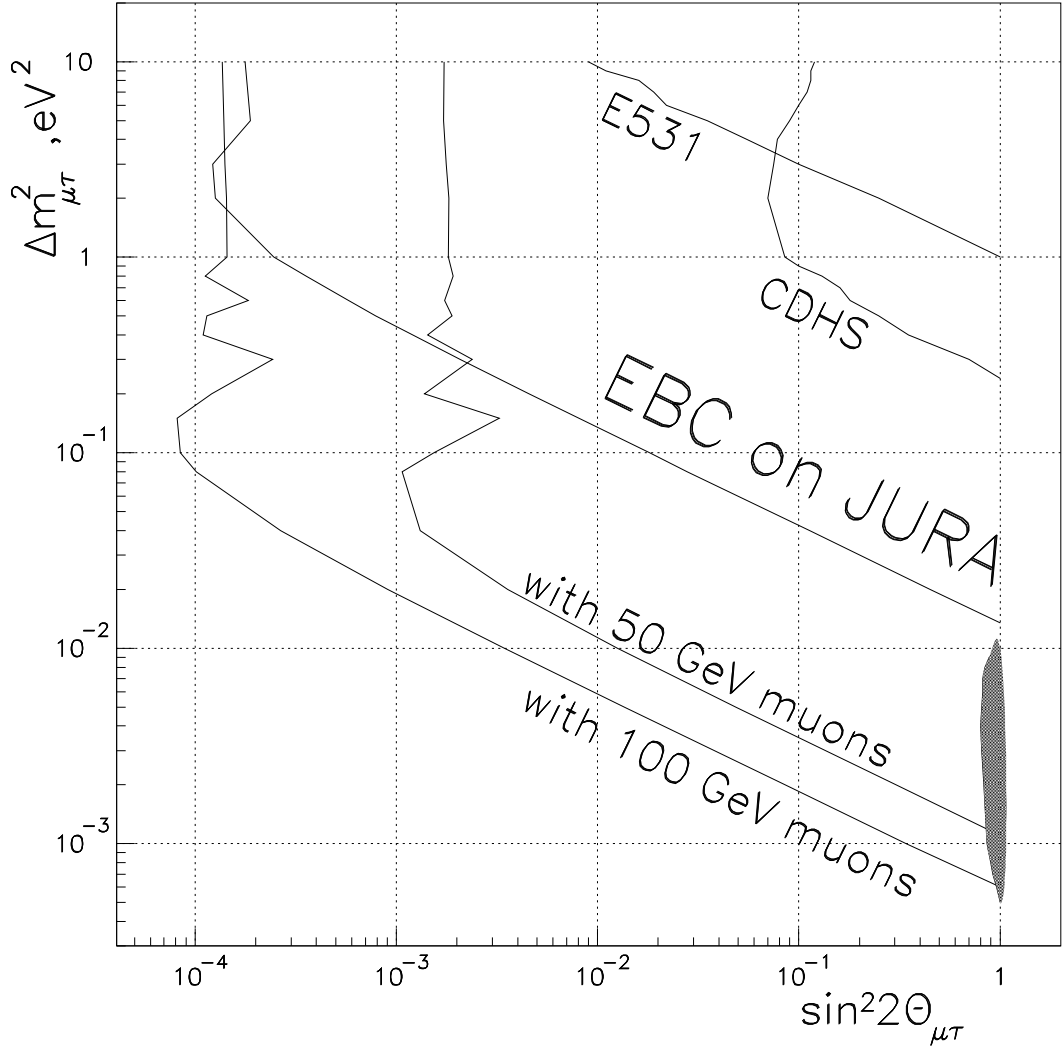} 
\caption{ 
Null-limit sensitivity to the \omutau\ transition (at 90\% C.L.) of a
20-ton EBC detector deployed at 17 km from CERN-SPS and at 732 km from a 
muon storage ring with $E_\mu = 50$ and 100 GeV, assuming 10$^{20}$ protons 
delivered
by CERN-SPS and  $2.2 \times 10^{21}$ negative muons injected in a ring with
a straight section of 25\%. The shaded area on the right is the region of
parameter space for \omutau\ suggested by a combined analysis of Kamiokande
and Superkamiokande data \cite{gonzalez}. Also illustrated are the best
upper limits on $\sin^2 2\theta_{\mu\tau}$  for  $\Delta m^2_{\mu\tau} < 10$
eV$^2$, as imposed by E531 \cite{e531} and CDHS \cite{cdhs} 
.}
\label{exclude}
\end{figure}

\noindent Fig. \ref{exclude}. \\
Null-limit sensitivity to the \omutau\ transition (at 90\% C.L.) of a
20-ton EBC detector deployed at 17 km from CERN-SPS and at
732 km from a muon storage ring with $E_\mu = 50$ and 100 GeV, assuming 
10$^{20}$ protons delivered
by CERN-SPS and  $2.2 \times 10^{21}$ negative muons injected in a ring with
a straight section of 25\%. The shaded area on the right is the region of
parameter space for \omutau\ suggested by a combined analysis of Kamiokande
and Superkamiokande data \cite{gonzalez}. Also illustrated are the best
upper limits on  $\sin^2 2\theta_{\mu\tau}$  for  $\Delta m^2_{\mu\tau} < 10$
eV$^2$, as imposed by E531 \cite{e531} and CDHS \cite{cdhs}.

\clearpage

\end{document}